\documentclass[12,epsf]{article}
\usepackage{graphicx}

\addtocounter{page}{1} \tolerance=5000
\newcommand{\beq}{\begin{equation}}
\newcommand{\eeq}{\end{equation}}
\newcommand{\beqn}{\begin{eqnarray}}
\newcommand{\eeqn}{\end{eqnarray}}

\newcommand{\lsim}{\mbox{$<$\hspace{-0.8em}\raisebox{-0.4em}{$\sim$}}}

\newcommand{\bJ}{\mathbf{J}}
\newcommand{\bj}{\mathbf{j}}

\newcommand{\al}{\mbox{${\alpha}$}}

\begin{document}

\begin{titlepage}

\begin{center}

{\Large \bf  HOW ARE BLACK HOLES QUANTIZED?}

\vspace{9mm}

{\bf I.B. Khriplovich}\footnote{khriplovich@inp.nsk.su}

\vspace{9mm}

{\em Budker Institute of Nuclear Physics, 630090 Novosibirsk,
Russia,\\ and Novosibirsk University}

\end{center}

\vspace{9mm}

\begin{abstract}

Some approaches to quantization of the horizon area of black holes
are discussed. The maximum entropy of a quantized surface is
demonstrated to be proportional to the surface area in the
classical limit. This result is valid for a rather general class
of approaches to surface quantization. In the case of rotating
black holes no satisfactory solution for the quantization problem
has been found up to now.

\end{abstract}

\vspace{10cm}

\end{titlepage}

\section{Introduction}

Some properties of black holes, both classical and quantum, were
discussed at XXXII Winter School of PNPI \cite{kh1}. The last part
of that talk referred to the quantization of black holes. Here we
will consider this problem from somewhat different point of view.
But for the completeness sake let us recall at first some
considerations from \cite{kh1}.

The idea of quantizing the horizon area of black holes was proposed by
Bekenstein many years ago \cite{bek}. It was based on the fact that the
horizon area of a nonextremal black hole behaves in a sense as an adiabatic
invariant \cite{chr}. And the quantization of an adiabatic invariant looks
perfectly natural. Once this hypothesis is accepted, the general structure
of the quantization condition for large quantum numbers gets obvious up to
an overall numerical constant $\al$ (our argument here goes
back to \cite{kh2} and somewhat differs from that of the original one \cite{bek}).
The quantization condition for the horizon area $A$ should be
\begin{equation}\label{qu}
A=\alpha \, l^2_p \, N,
\end{equation}
where $N$ is some large quantum number which generally speaking should not
be an integer. Indeed, the presence of the Planck length squared
\begin{equation}
l^2_p = k \hbar /c^3
\end{equation}
(here $k$ is the Newton gravitational constant) is only natural in this
quantization rule. Then, for the horizon area $A$ to be finite in the
classical limit, the power of $N$ here should be the same as that of $\hbar$
in $l^2_p$.

From different points of view the black hole quantization was discussed
later in Refs. \cite{muk,kog}. In particular, the article by Kogan \cite{kog}
was the first investigation of the problem in the frame of the very popular
now string approach. However, though a lot of work has been done since on
the subject, especially based on the string theory and on the loop quantum
gravity, I do not think that the problem is really solved now. Some attempts
to attack the problem of black hole quantization are described below.

\section{"It from bit"}

The quantization condition (\ref{qu}) suggests that the horizon surface
splits into patches of the area $\sim l^2_p$ each. But what can one say
about the contribution by each of them into the total area?

Perhaps, the simplest model of the quantization, naturally called "it from
bit", is due to Wheeler \cite{whe}. Here the horizon surface consists of $\nu$
patches of area $\sim l^2_p$ (of course, for a classical black hole $\nu
\gg 1$), and to each of them one assigns a spin variable with two possible
values $j_z= +1/2$ and $j_z= -1/2$. Thus, the total number of states is $K =
2^{\nu}$. The black hole entropy defined as log of number of states is
\begin{equation}\label{sn}
S= \ln K = \nu \,\ln 2 .
\end{equation}
On the other hand, the entropy of a black hole is related to the area $A$ of
its horizon by the famous Bekenstein-Hawking relation (see, e.g., \cite{kh1})
\begin{equation}\label{bh}
A = 4 \, l^2_p \, S.
\end{equation}
In this way one arrives at the quantization rule for the horizon area
\begin{equation}
A = 4 \ln2\, l^2_p \,\nu,
\end{equation}
with an integer $\nu$ and fixed overall numerical constant $\al = 4 \ln2$.

It is absolutely crucial here that the patches should be
distinguishable. Otherwise, instead of $2^{\nu}$ states, we would
have only $\nu+1$ states: a single one with all spins up, a
single one with one spin down and all other spins up, a single
one with two spins down and all other spins up, and so on. Thus,
with indistinguishable patches the entropy would be $S =
\ln(\nu+1)$, which has nothing in common with the required
proportionality to the area $A \sim l^2_p \, \nu$, especially if
one recalls the classical condition $\nu \gg 1$.

One more point should be mentioned here. With the spins assigned
to the patches being the only building blocks of the model, it
will be natural to assume that the total angular momentum $J$ of
the black hole is just the sum of these spins. It is clear
intuitively that for the random distribution of $j_{z}$ over the
patches this total angular momentum should be small. To find how
small exactly, let us calculate at first the number $K(J_{z}=0)$
of states with $J_{z}=\Sigma j_{z}=0$. Evidently, here the number
$\nu{\pm}$ of patches with spins up and down is
\[
\nu_+ =\nu_- = \nu / 2 \quad (\nu \gg 1).
\]
Thus,
\[
K(J_{z}=0)={\nu ! \over [(\nu /2)!]^2}\,.
\]
And the entropy of this state, as estimated for $\nu \gg 1$ with the
Stirling formula,
\begin{equation}\label{sn0}
S(J_{z}=0)=\ln K(J_{z}=0)= \nu \,\ln 2 - {1 \over 2}\ln \nu
\end{equation}
indeed differs from the initial one (\ref{sn}) by a logarithmic correction
only.

As to the number of states with $J=0$, $K(J=0)$, it equals obviously to the
difference between the numbers of the states with $J_{z}=0$ and $J_{z}=\pm 1$:
\[
K(J=0)=K(J_{z}=0)-K(J_{z}=\pm 1)={\nu ! \over [(\nu
/2)!]^{2}}-{\nu ! \over (\nu /2-1)!(\nu /2+1)!}={\nu !\over (\nu
/2)!(\nu /2+1)!}\,.
\]
The entropy of this state, with $J=0$,
\begin{equation}\label{sn00}
S(J=0)=\ln K(J=0)= \nu \,\ln 2 -{3 \over 2}\ln \nu
\end{equation}
again differs from (\ref{sn}) (and from (\ref{sn0}) as well) by a
logarithmic correction only.

Thus, even without the subsidiary condition $J = 0$, the simple-minded "it
from bit" model can describe in a rather natural way the Schwarzschild black
hole.

Let us go over now to the Kerr black hole, rotating one. Its entropy looks
as follows (see, e.g., \cite{kh1}):
\begin{equation}\label{sk}
S = 2\pi \left(M^2 + \sqrt{M^4 - J^2}\,\right);
\end{equation}
here $M$ is the black hole mass (to simplify formulae we put now $c=1$,
$\hbar=1$, $k=1$).
As to the black hole angular momentum $J$, we assume again that it is the sum
of the spins of the patches: $J=\Sigma \bj$. Let us see whether the
"it from bit" model can reproduce in this way the semiclassical relation
(\ref{sk}). It can be easily demonstrated that in the discussed model
the number of states with total angular momentum $J$ equals
\[
K(J) = (2J+1)\,[\,K(J_{z}=J)-K(J_{z}=J+1)\,]
\]
\beq
=\nu!(2J+1)\left[{1 \over (\nu /2-J)!(\nu /2+J)!} - {1 \over (\nu
/2-J-1)!(\nu /2+J+1)!}\right]
\eeq
\[
= {\nu !(2J+1)^{2} \over (\nu /2-J)!(\nu /2+J+1)!}
\]
In the
limiting case $1\ll J\ll \nu $ the "it from bit" entropy is
\beq\label{ifb}
S_{{\rm ifb}}=\ln K(J)= \nu \, \ln 2 - {2J^{2} \over \nu }\,.
\eeq
But let us look now at the corresponding expansion of the semiclassical
formula (\ref{sk}):
\beq\label{sk1}
S=2\pi \left(2M^2 - {J^2 \over 2M^{2}}\,\right).
\eeq
Obviously, we should identify $2M^{2}$ with $\nu \ln 2/2\pi $.
But then formula (\ref{sk1}) is rewritten as
\beq\label{sk2}
S=\nu \, \ln 2 - {(2\pi )^{2} \over \ln 2}\,{J^{2} \over \nu }\,.
\eeq
The difference between numerical factors at $J^{2}/\nu $ is tremendous,
the "it from bit" model strongly underestimates the entropy decrease
with $J$ in the discussed regime.

In the limit of the extremal black hole, where according to the
semiclassical formula (\ref{sk}) $J$ acquires its maximum value
$M^2$ and the entropy equals $2\pi M^2$, the situation with "it
from bit" is even worse. One option is to assume here for $J$ its
maximum possible value in this model which is $\nu/2$. In this
case the number of states is at most $2J+1= \nu + 1$ (if one
counts possible orientations of $\mathbf{J}$), and the entropy
$\ln(\nu + 1)$ is negligibly small as compared to its
semiclassical value $2\pi M^2 = \nu \ln 2/2$. Another option is
to assign here to $J$ its semiclassical limiting value $M^2 = \nu
\ln 2/4\pi$. But this value is much smaller numerically than
$\nu$, and in this way we effectively arrive again at formula
(\ref{ifb}), so that the extremal entropy becomes very close
numerically to the Schwarzschild one instead of constituting half
of it. Thus, in general case {\em bit is insufficient for it}.

\section{Loop quantum gravity\\ and maximum entropy principle}

In a sense, the result for the quantized horizon surface derived
in loop quantum gravity [8--12] can be considered as a
generalization of the "it from bit" model. Here, a surface
geometry is determined by a set of $\nu$ punctures on this
surface (recall the patches in the "it from bit" picture). In
general, each puncture is supplied by two integer or half-integer
angular momenta $j^u$ and $j^d$:
\begin{equation}
j^u, j^d = 0, 1/2, 1, 3/2, ...
\end{equation}
$j^u$ and $j^d$ are related to edges directed up and down the normal to the
surface, respectively, and add up into an angular momentum $j^{ud}$:
\begin{equation}
\mathbf{j}^{ud} = \mathbf{j}^{u} + \mathbf{j}^{d}; \quad |j^u-j^d|\leq
j^{ud}\leq |j^u+j^d|.
\end{equation}
The area of a surface is
\begin{equation} \label{Aq}
A = \al \, l^2_p \, \sum^{\nu}_{i=1}\sqrt{2j^u(j^u+1) +
2j^d(j^d+1) - j^{ud}(j^{ud}+1)}\,.
\end{equation}
A comment on the last formula is appropriate. Since the quantum numbers $j$
entering it can be in principle arbitralily large, the same correspondence
between the power of a quantum number and the power of $\hbar$ (hidden here
in $l^2_p$) in the classical limit, dictates that just the sum of square
roots, but not for instance the sum of $j(j+1)$ should enter the expression
for the area.

As to the overall numerical factor $\al$ in (\ref{Aq}), it
cannot be determined without an additional physical input. This ambiguity
originates from a free (so-called Immirzi) parameter \cite{imm,thi} which
corresponds to a family of inequivalent quantum theories, all of them being
viable without such an input.

In the case of a Schwarzschild black hole, the problem was
attacked in \cite{asht} by introducing into the theory boundary
terms corresponding to a Chern-Simons theory on the horizon. The
approach results in the situation when a single spin j=1/2 with
two possible projections is attached to each puncture. In other
words, here each radical in formula (\ref{Aq}) contains only
$j^{u(d)}=j$, $j^{d(u)}=0$, $j^{ud}=j$, and thus reduces to
$\sqrt{j(j+1)}=\sqrt{3}/2$. Obviously, in this case the entropy
equals again $\nu \ln2$, and with the Bekenstein-Hawking relation
(\ref{bh}) one fixes immediately the numerical parameter $\al$ in
formula (\ref{Aq}): $\al = 2 \ln2/\sqrt{3}$. Further elaborations
on this approach \cite{km,car} (see also \cite{gour}) resulted in
more accurate formula (\ref{sn00}) for the entropy of a
nonrotating black hole: $S = \nu\,\ln 2 - 3/2\ln \nu$. Thus, for
the Schwarzschild black hole, one effectively arrives here again
at the "it from bit" picture with all partial spins added up into
$J = 0$. However, as to the Kerr black hole, we know already that
{\em bit is insufficient for it}.

It is natural to investigate now whether the general Ansatz (\ref{Aq}) is
more appropriate
for the description of black holes. As to the value of the overall numerical
factor $\al$ in (\ref{Aq}), one may hope that it can be determined
by studying the entropy of a black hole. It has been done indeed in \cite{kh3,kk}
under the assumption that the entropy of an eternal black hole in equilibrium
is maximum. This assumption goes back to \cite{vaz}, where it was used in a
model of the quantum black hole as originating from dust collapse. It looks quite
natural from physical point of view to assume that among the entropies of various
surfaces of the given area, it is the entropy of the black hole horizon which is
the maximum one.

The entropy of a surface is defined as the logarithm of the number of states of
this surface with fixed area, i. e.  fixed sum (\ref{Aq}). Let $\nu_i$ be the
number of punctures with a given set of momenta $j^{u}_i$, $j^{d}_i$, $j^{ud}_i$.
The total number of punctures is
\[
\nu = \sum_i \nu_i.
\]
To each puncture $i$ one ascribes a statistical weight $g_i$. Since
$\mathbf{j}^{ud} = \mathbf{j}^{u} + \mathbf{j}^{d}$, this statistical weight equals,
in the absence of other constraints, to the number of possible projections of
$j^{ud}_i$, i. e. $g_i = 2 j^{ud}_i + 1$. Then the entropy is
\beq\label{Sq}
S = \ln\left[\prod_i {(g_i)^{\nu_i} \over \nu_i !} \nu ! \right].
\eeq
The structure of expressions (\ref{Aq}) and (\ref{Sq}) is so different that the
proportionality between them for general values of $j^{u}_i$, $j^{d}_i$, $j^{ud}_i$
may look impossible. However, as will be demonstrated now, this is the case for
the maximum entropy in the classical limit.

But first of all let us note that the requirement of the entropy being
proportional to the area is indeed quite nontrivial and restrictive. First of all,
it excludes the presence of "empty" punctures, with $j^{u}_i = j^{d}_i = 0$. Such
punctures would not influence the area $A$, would increase the combinatorics, and
with it the entropy $S$. These "empty" punctures are excluded indeed in loop quantum
gravity, but by quite different arguments.

On the other hand, this requirement forbids an excessive
concentration of angular momenta at some relatively small number
of punctures. The limiting case of such a concentration takes
place in the black hole model suggested in \cite{cas}. Therein
all angular momenta are collected effectively into a single one
$J \gg 1$ at a single puncture, with the statistical weight
$2J+1$. Thus, while the area in the model \cite{cas} is $A \sim
\sqrt{J(J+1)} \sim J$, the entropy is $S \sim \ln(2J+1) \sim \ln
J$.

By combinatorial reasons, it is natural to expect that the absolute maximum of
entropy is reached when all values of quantum numbers $j^{u,d,ud}_i$ are present.
This guess is confirmed by concrete calculations for some model cases \cite{kh3}.
We assume also that in the classical limit the typical values of puncture numbers
$\nu_i$ are large. Then, with the Stirling formula for factorials, formula (\ref{Sq})
transforms to
\beq\label{Sc}
S = \sum_i \left[\nu_i \ln g_i - \left(\nu_i + {1 \over 2}\right)
\ln \nu_i \right] + \left(\sum_i \nu_i + {1 \over 2}\right)
\times  \ln \left(\sum_{i'}\nu_{i'}\right).
\eeq
We have omitted here terms with $\ln\sqrt{2\pi}$, each of them being on the order
of unity. The validity of this approximation, as well as of the Stirling formula
by itself for this problem, will be discussed later.

We are looking for the extremum of expression (\ref{Sc}) under the condition
\beq\label{con}
N = \sum_i \nu_i r_i = {\rm const}, \quad r_i = \sqrt{2j^u(j^u+1)
+ 2j^d(j^d+1) - j^{ud}(j^{ud}+1)}\,.
\eeq
The problem reduces to the solution of the system of equations
\beq\label{eq}
\ln g_i - \ln \nu_i + \left(\sum_{i'} \ln \nu_{i'}\right) = \mu
r_i\, ,
\eeq
or
\beq\label{eq1}
\nu_i = g_i \exp(- \mu r_i)\sum_{i'} \nu_{i'}\, .
\eeq
Here $\mu$ is the Lagrange multiplier for the constraining relation (\ref{con}).
Summing expressions (\ref{eq1}) over $i$, we arrive at the equation on $\mu$:
\beq\label{sec}
\sum_i g_i \exp(- \mu r_i) = 1.
\eeq
On the other hand, when multiplying equation (\ref{eq}) by $\nu_i$ and summing
over $i$, we arrive with the constraint (\ref{con}) at the following result for
the maximum entropy for a given value of $N$:
\beq\label{Sm}
S_{{\rm max}} = \mu N\,.
\eeq
Here the terms
\[
- {1 \over 2} \sum_i \ln \nu_i \quad {\rm and} \quad {1 \over
2}\ln \left(\sum_{i'}\nu_{i'}\right)
\]
in expression (\ref{Sc}) have been neglected. Below we will come back to the
accuracy of this approximation.

Thus, it is the maximum entropy of a quantized surface which is proportional
in the classical limit to its area. This proportionality certainly exists
for a classical black hole. And this is a very strong argument in favour of
the assumption that the black hole entropy is maximum.

It should be stressed that relation (\ref{Sm}) is true not only in
the loop quantum gravity, but applies to a more general class of
approaches to the quantization of surfaces. What is really
necessary here, is as follows. The surface should consist of
patches of different sorts, so that there are $\nu_i$ patches of
a given sort $i$, each of them possessing a generalized effective
quantum number $r_i$ and a statistical weight $g_i$. Then in the
classical limit, the maximum entropy of a surface is proportional
to its area.

Let us consider the general case, with $N$ given by formula (\ref{con}),
$g_i = 2 j^{ud}_i + 1$, and all values of $j^{u}_i$, $j^{d}_i$, $j^{ud}_i$
allowed. Here the numerical solution of equation (\ref{con}) is
$\mu = 3.120$, and the maximum entropy equals
\beq
S = 3.120 N = 4.836 \nu.
\eeq
The mean values of quantum numbers are
\beq
\langle j^{u}_i \rangle = \langle j^{d}_i \rangle = 1.072, \quad
\langle j^{ud}_i \rangle  = 2.129.
\eeq
Again, it is clear intuitively, and demonstrated explicitly by the
above analysis for the simpler case of "it from bit" model, that
with the assumed random distribution of $j^{ud}_{z}$ over the
patches the total angular momentum $\bJ = \sum \bj^{ud}_i$ should
effectively vanish. Thus, the considered model describes a
Schwarzschild black hole.

It should be stressed that in this way one always arrives at the
quantization rule for the black hole entropy and area effectively
with integer quantum numbers $\nu$, as was proposed in the
pioneering article \cite{bek}.

Let us discuss now the accuracy of our result for the maximum
entropy. With all $\langle j \rangle \simeq 1$, the number of
punctures $\nu$ is on the same order of magnitude as $N$. Thus,
in the classical limit, $\nu \sim N \gg 1$. Now, according to
relation (\ref{eq1}), with $\exp(- \mu) \lsim 1$, the numbers
$\nu_i$ satisfy the condition $\nu_i > 1$ as long as the quantum
numbers $j$ are bounded by condition
\beq\label{in}
j \lsim \ln N.
\eeq
Clearly, the typical values of those $\nu_i$ which contribute essentially
to $N$ are large, and the Stirling approximation for $S$ is fully
legitimate.

On the other hand, the number of terms in the sums in expression
(\ref{Sc}) is effectively bounded by inequality (\ref{in}). Thus,
the contribution of the terms with $\ln \sqrt{2\pi}$, omitted in
(\ref{Sc}), as well as of the term $1/2\; \ln \nu$ retained
therein, but neglected in the final expression (\ref{Sm}), is on
the order of $\ln N$ only. The leading correction to our result
(\ref{Sm}) originates from the term $- 1/2\; \sum \ln \nu_i$, and
constitutes $\sim \ln^2 N$ in order of magnitude.

Few words on the attempts made in \cite{rove,kra} to calculate the
surface entropy in loop quantum gravity. In those
papers the distribution of the angular momenta $j$ over the punctures is
not discussed at all. We cannot understand how one could find the surface
entropy without such information.

But what about the description of Kerr black holes in the discussed
approach? It looks reasonable here to look for the maximum entropy by
introducing, in line with the constraint (\ref{con}) of fixed area,
one more constraint, of fixed total angular momentum:
\beq\label{con1}
\sum_i j^{ud}_{z,i} = \sum_i (j^{u}_{z,i} + j^{d}_{z,i}) = J.
\eeq
As to the $x$- and $y$- projections of the angular momenta, with their
random distribution over the punctures, they effectively add up into
zero (see the analogous discussion in the "it from bit" section).

Some progress is achieved in this way for an extremal black hole:
its entropy does not vanish. However, instead of being two times
smaller than the entropy of a Schwarzschild black hole of the same
mass, it is smaller than the last one by a factor about 1.5 only.
As to the entropy decrease for small $J$, it is far too small (as
this was the case in the "it from bit" model). So much the more,
we have not succeeded in reproducing the general
relation~(\ref{sk}).

Up to now no satisfactory solution to the problem of description
of Kerr black holes in loop quantum gravity has been pointed out.
This is a real challenge for the theory.

\vspace{5mm}

\noindent{\bf Acknowledgents}

\vspace{3mm}

\noindent I am grateful to R.V. Korkin, the content of the talk is
essentially based on the work done together with him. The
investigation was supported in part by the Russian Foundation for
Basic Research through Grant No. 01-02-16898, through Grant No.
00-15-96811 for Leading Scientific Schools, by the Ministry of
Education Grant No. E00-3.3-148, and by the Federal Program
Integration-2001.

\end{document}